\documentclass[amsmath,amssymb,superscriptaddress,nobalancelastpage,prb,twocolumn]{revtex4-1}

\usepackage{physics, amsmath,amssymb, siunitx}
\usepackage{graphicx}
\usepackage{booktabs}
\usepackage{color}
\usepackage{xcolor}
\usepackage{enumitem}
\usepackage[normalem]{ulem}
\setlist{noitemsep,leftmargin=*}
\usepackage{hyperref}
\hypersetup{
    colorlinks=true,
    urlcolor= blue,
    citecolor=blue,
    linkcolor= blue,
    bookmarksopen=false,
    }
\usepackage[titletoc,toc,title]{appendix}

\newcommand{\PFATO}{Pr$_{4}$Fe$_{2}$As$_2$Te$_{0.88}$O$_4$}

\newcommand{\dxy}{$d_{xy}$}
\newcommand{\dxz}{$d_{xz}$}
\newcommand{\dyz}{$d_{yz}$}
\newcommand{\EF}{$E_\mathrm{F}$}
\newcommand{\Tc}{$T_\mathrm{c}$}
\newcommand{\Tort}{$T_\mathrm{ort}$}

\newcounter{para}

\begin{document}

\title{Decoupling of Lattice and Orbital Degrees of Freedom in an Iron-Pnictide Superconductor
}

\author{C. E.~Matt}
\affiliation{Swiss Light Source, Paul Scherrer Institut, CH-5232 Villigen PSI, Switzerland}
\affiliation{Physik-Institut, Universit\"{a}t Z\"{u}rich, Winterthurerstrasse 190, CH-8057 Z\"{u}rich, Switzerland}

\author{O. Ivashko}
\affiliation{Physik-Institut, Universit\"{a}t Z\"{u}rich, Winterthurerstrasse 190, CH-8057 Z\"{u}rich, Switzerland}

\author{M. Horio}
\affiliation{Physik-Institut, Universit\"{a}t Z\"{u}rich, Winterthurerstrasse 190, CH-8057 Z\"{u}rich, Switzerland}

\author{D.~Sutter}
\affiliation{Physik-Institut, Universit\"{a}t Z\"{u}rich, Winterthurerstrasse 190, CH-8057 Z\"{u}rich, Switzerland}

\author{N. Dennler}
\affiliation{Physik-Institut, Universit\"{a}t Z\"{u}rich, Winterthurerstrasse 190, CH-8057 Z\"{u}rich, Switzerland}

\author{J. Choi}
\affiliation{Physik-Institut, Universit\"{a}t Z\"{u}rich, Winterthurerstrasse 190, CH-8057 Z\"{u}rich, Switzerland}

\author{Q. Wang}
\affiliation{Physik-Institut, Universit\"{a}t Z\"{u}rich, Winterthurerstrasse 190, CH-8057 Z\"{u}rich, Switzerland}

\author{M.~H.~Fischer}
\affiliation{Physik-Institut, Universit\"{a}t Z\"{u}rich, Winterthurerstrasse 190, CH-8057 Z\"{u}rich, Switzerland}


\author{S. Katrych}
\affiliation{Laboratory of Physics of Complex Matter, \'Ecole Polytechnique F\'ed\'erale de Lausanne (EPFL), CH-1015 Lausanne, Switzerland}

\author{L. Forro}
\affiliation{Laboratory of Physics of Complex Matter, \'Ecole Polytechnique F\'ed\'erale de Lausanne (EPFL), CH-1015 Lausanne, Switzerland}

\author{J. Ma}
\affiliation{Swiss Light Source, Paul Scherrer Institut, CH-5232 Villigen PSI, Switzerland} 
\author{B. Fu}
\affiliation{Swiss Light Source, Paul Scherrer Institut, CH-5232 Villigen PSI, Switzerland} 
\author{B. Lv}
\affiliation{Swiss Light Source, Paul Scherrer Institut, CH-5232 Villigen PSI, Switzerland} 

\author{M.~v.~Zimmermann}
\affiliation{Deutsches Elektronen-Synchrotron DESY, 22603 Hamburg, Germany}

\author{T. K. Kim}
\affiliation{Diamond Light Source, Harwell Campus, Didcot, OX11 0DE, United Kingdom}

\author{N.~C.~Plumb}
\affiliation{Swiss Light Source, Paul Scherrer Institut, CH-5232 Villigen PSI, Switzerland}

\author{N. Xu}
\affiliation{Swiss Light Source, Paul Scherrer Institut, CH-5232 Villigen PSI, Switzerland}


\author{M.~Shi}
\affiliation{Swiss Light Source, Paul Scherrer Institut, CH-5232 Villigen PSI, Switzerland}  

\author{J.~Chang}
\affiliation{Physik-Institut, Universit\"{a}t Z\"{u}rich, Winterthurerstrasse 190, CH-8057 Z\"{u}rich, Switzerland}

\maketitle

\textbf{The interplay of structural and electronic phases in iron-based superconductors is a central theme in the search for the superconducting pairing mechanism. While electronic nematicity, defined as the breaking of four-fold symmetry triggered by electronic degrees of freedom, is competing with superconductivity \cite{RMFernandesNPHYS2014, NandiPRL2010}, the effect of purely structural orthorhombic order is unexplored. Here, using x-ray diffraction (XRD), we reveal a new structural orthorhombic phase with an exceptionally high onset temperature ($T_\mathrm{ort} \sim 250$ K), which coexists with superconductivity ($T_\mathrm{c} = 25$ K), in an electron-doped iron-pnictide superconductor far from the underdoped region. Furthermore, our angle-resolved photoemission spectroscopy (ARPES) measurements demonstrate the absence of electronic nematic order as the driving mechanism, in contrast to other underdoped iron pnictides where nematicity is commonly found. Our results establish a new, high temperature phase in the phase diagram of iron-pnictide superconductors and impose strong constraints for the modeling of their superconducting pairing mechanism.}

\begin{figure*}
 	\begin{center}
 		\includegraphics[width=0.99\textwidth]{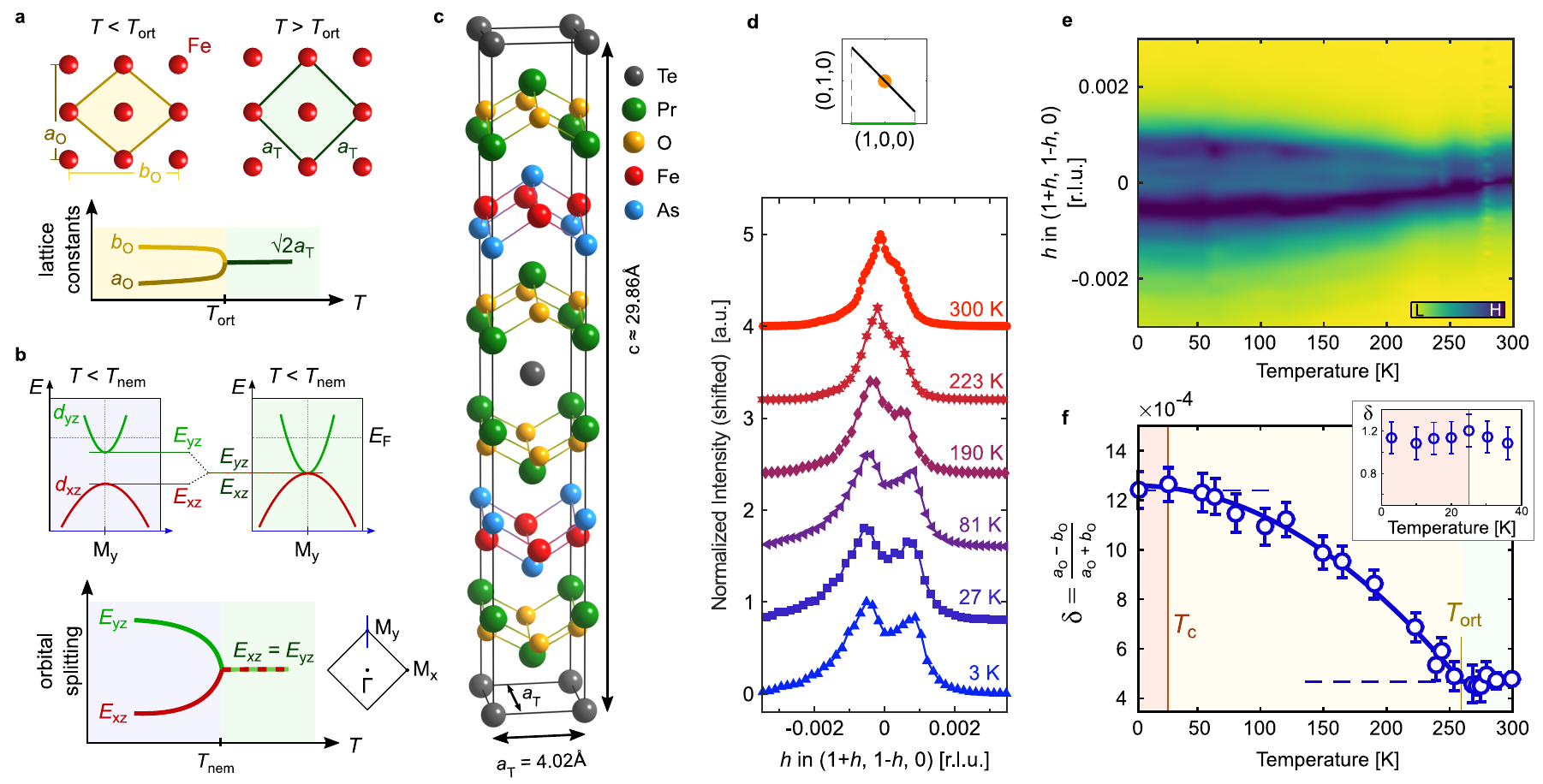}
 	\end{center}
 	\caption{\textbf{High-temperature orthorhombicity in Pr$_{4}$Fe$_{2}$As$_2$Te$_{0.88}$O$_4$ (PFATO).} \textbf{a,} Schematics for structural C$_4$ symmetry breaking at $T<T_\mathrm{ort}$ in iron pnictides. The order parameter is given by $\delta=(a_\mathrm{O}-b_\mathrm{O})/(a_\mathrm{O}+b_\mathrm{O})$, where $a_\mathrm{O}$ and $b_\mathrm{O}$ are the orthorhombic lattice parameters. 	\textbf{b,}  Schematics of orbital splitting in the nematic phase of underdoped iron-pnictides. Below $T_\mathrm{nem}$ the \dxz\ and \dyz\ are no longer degenerate and split in energy, with largest splitting at the M points of the Brillouin zone \cite{yinpjQuantMater2017}.  	Here, $E_\mathrm{xz}$ ($E_\mathrm{yz}$) denote the maximum (minimum) of the hole-like (electron-like) bands at M$_\mathrm{y}$. 
 	\textbf{c,} Room temperature  unit cell of Pr$_{4}$Fe$_{2}$As$_2$Te$_{0.88}$O$_4$ (space group \textit{I4/mmm}, 139).
\textbf{d-e,} High resolution transverse scans through the (1,1,0) Bragg  reflection (see inset) at temperatures as indicated.  A splitting of the diffraction peak is observed below $T_\mathrm{ort} \sim 250$ K.  \textbf{f,} Temperature dependence of the orthorhombic order parameter $\delta$. Inset:  $\delta$ around \Tc\ measured in finer temperature steps. }
	\label{fig:fig1}
 \end{figure*}

Underdoped iron-pnictides commonly host a nematic state below $T_\mathrm{nem}$ in which the four-fold rotational C$_4$ symmetry is spontaneously broken into C$_2$ symmetry in electron, spin and structural degrees of freedom\cite{GRStewartRMP2011} (see Figs. \ref{fig:fig1} a-b). Due to its close proximity to superconductivity and its putative quantum criticality\cite{kuoScience2016, ledererPhysRevLett2015, FradkinARCMP2010} the nematic phase has been heavily investigated for elucidating a microscopic description of the superconducting pairing mechanism. Among other experiments, transport \cite{ChuScience2010, chuScience2012a, TanatarPRB2010} and ARPES\cite{yiPNAS2011, kimPhysRevB2011a, yiPhysRevX2019, watsonPhysRevB2015, yiNewJPhys2012, zhangPhysRevB2012} studies provided compelling evidence that nematicity in underdoped iron-pnictides is triggered by electronic order, subsequently driving other degrees of freedom such as the lattice to break C$_4$ symmetry\cite{NandiPRL2010}. In ARPES the orbital order was directly revealed by the splitting of the  \dxz\ -- \dyz\ bands, which are degenerate in the high-temperature tetragonal phase (see Fig. \ref{fig:fig1} b). In some parent compounds, a band splitting of up to $\sim 60$ meV was observed\cite{yiPNAS2011, kimPhysRevB2011a,  watsonPhysRevB2015, yiPhysRevX2019}, which is too large to be a trivial consequence of the small orthorhombicity of less than $1$\% (ref. \onlinecite{huangPRL2008}). This led to the conclusion that electronic order is driving the nematic state in underdoped pnictides\cite{yinpjQuantMater2017}, triggering a tetragonal-to-orthorhombic transition in the lattice degrees of freedom at the same  temperature scale ($T_\mathrm{ort} = T_\mathrm{nem}$). Importantly, XRD measurements of the orthorhombic distortion revealed that the electronically driven nematic phase is competing with superconductivity\cite{NandiPRL2010}. However, due to the strong coupling of electronic and lattice degrees of freedom, the competing channel (i.e. electronic or structural) is ambiguous. 


 \begin{figure*}
 	\begin{center}
 		\includegraphics[width=0.99\textwidth]{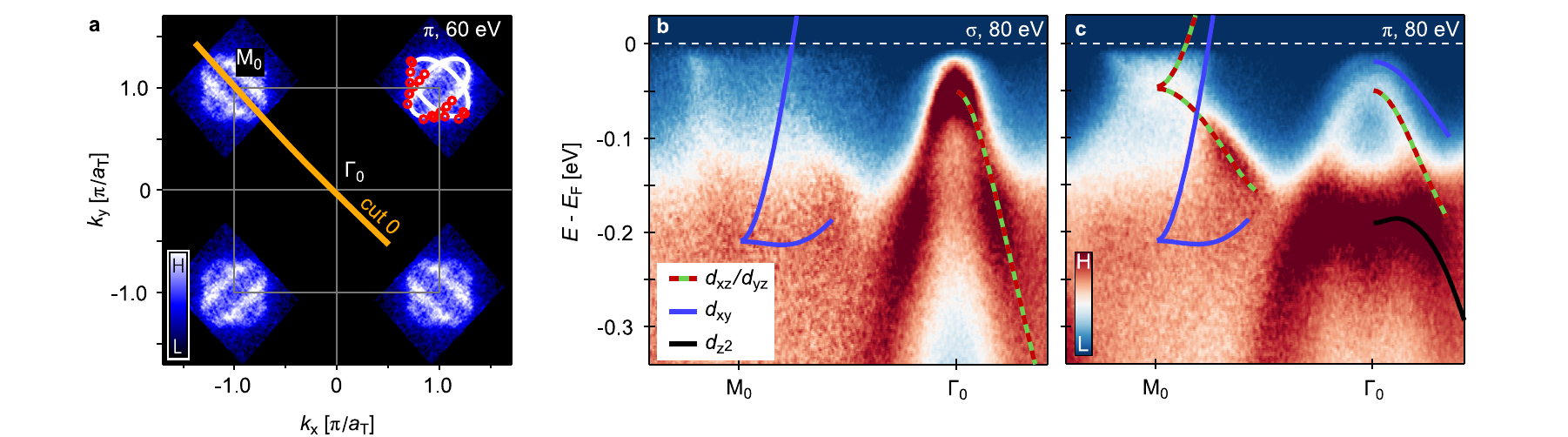}
 	\end{center}
 	\caption{\textbf{Electronic band structure of \PFATO.}
        \textbf{a,} Symmetrised Fermi surface map recorded on PFATO around zone corners and displayed in false colour scale. The Fermi surface contours of the two electron pockets are shown schematically with solid lines in the top right zone corner. The two underlying Fermi pockets amount to a combined filling of ~0.11$\pm 0.01$ electrons per Fe (for more information, see the supplementary material).  \textbf{b-c,} Low-energy electronic band structure along a cut 0 ($\Gamma_0 - M_0$) measured with $\sigma$ and $\pi$ polarised light ($h\nu=80$~eV). Overlaid lines are shifted and renormalised DFT calculated bands. The colour code identifies the orbital  character of the bands. All spectra were recorded at 5.5 K.
  	 }
	\label{fig:fig2}
 \end{figure*}

Here, we report the observation of a new, purely structural orthorhombic phase in the electron-doped iron-pnictide superconductor  Pr$_{4}$Fe$_{2}$As$_2$Te$_{0.88}$O$_4$ (PFATO, $T_\mathrm{c}=25$~K)~\cite{KatrychPRB2014,KatrychPRB2013,PisoniPRB2016}. Our XRD study revealed that this phase has an exceptionally high onset temperature of $T_\mathrm{ort} = 250$ K. This is $\sim 50$ K higher than the maximum \Tort\ which has so far been observed in iron-based superconductors \cite{GRStewartRMP2011}. Considering the high electron doping of 0.12 electrons per Fe, this observation is unexpected. Furthermore, we find that the low-temperature structural orthorhombic order parameter $\delta$ is not suppressed below $T_\mathrm{c}$ indicating a "friendly" coexistence of orthorhombicity with superconductivity. The structural (phononic) origin of this phase is further supported by our ARPES measurements which reveal the absence of orbital splitting down to lowest temperatures within the applied energy resolution. These properties are orthogonal to what has been found for the nematic phase in underdoped iron-pnictides and therefore indicate that this orthorhombic phase is new and distinct from previously described phases in iron-pnictides. \\[2mm]

We directly reveal the tetragonal-to-orthorhombic phase transition in PFATO by XRD measurements of the (1,1,0) Bragg reflection. At room temperature, 
x-ray diffraction studies have previously shown that PFATO has a tetragonal \textit{I4/mmm} structure with a large $c$-axis (29.86~\AA) lattice parameter \cite{KatrychPRB2013} (see Fig.~\ref{fig:fig1} c). As shown in Fig. \ref{fig:fig1} d-e, below $T_\mathrm{ort}=250$~K, the (1,1,0) reflection (inset of Fig.~\ref{fig:fig1} d) splits into two peaks along the transverse direction, evidencing the existence of orthorhombicity.  The gradual (continuous) onset of the orthorhombic order parameter at \Tort\ suggests a second-order phase transition\cite{TonegawaNatCom2014}, congruent to what is found in parent compounds of iron-pnictides \cite{GRStewartRMP2011} and FeSe\cite{ABoehmerJPCM2018}.
However, the onset temperature is $\sim 50$ K higher than the highest structural transition temperature in any other iron-pnictide parent compound \cite{GRStewartRMP2011}. In light of the rather high electron doping of \PFATO\ (0.12 electrons per Fe, see supplementary information), such a high onset temperature is surprising and indicates that this orthorhombic phase is different in nature from the electronic nematic phase in underdoped iron-pnictides.

\begin{figure*}
 	\begin{center}
 		\includegraphics[width=0.99\textwidth]{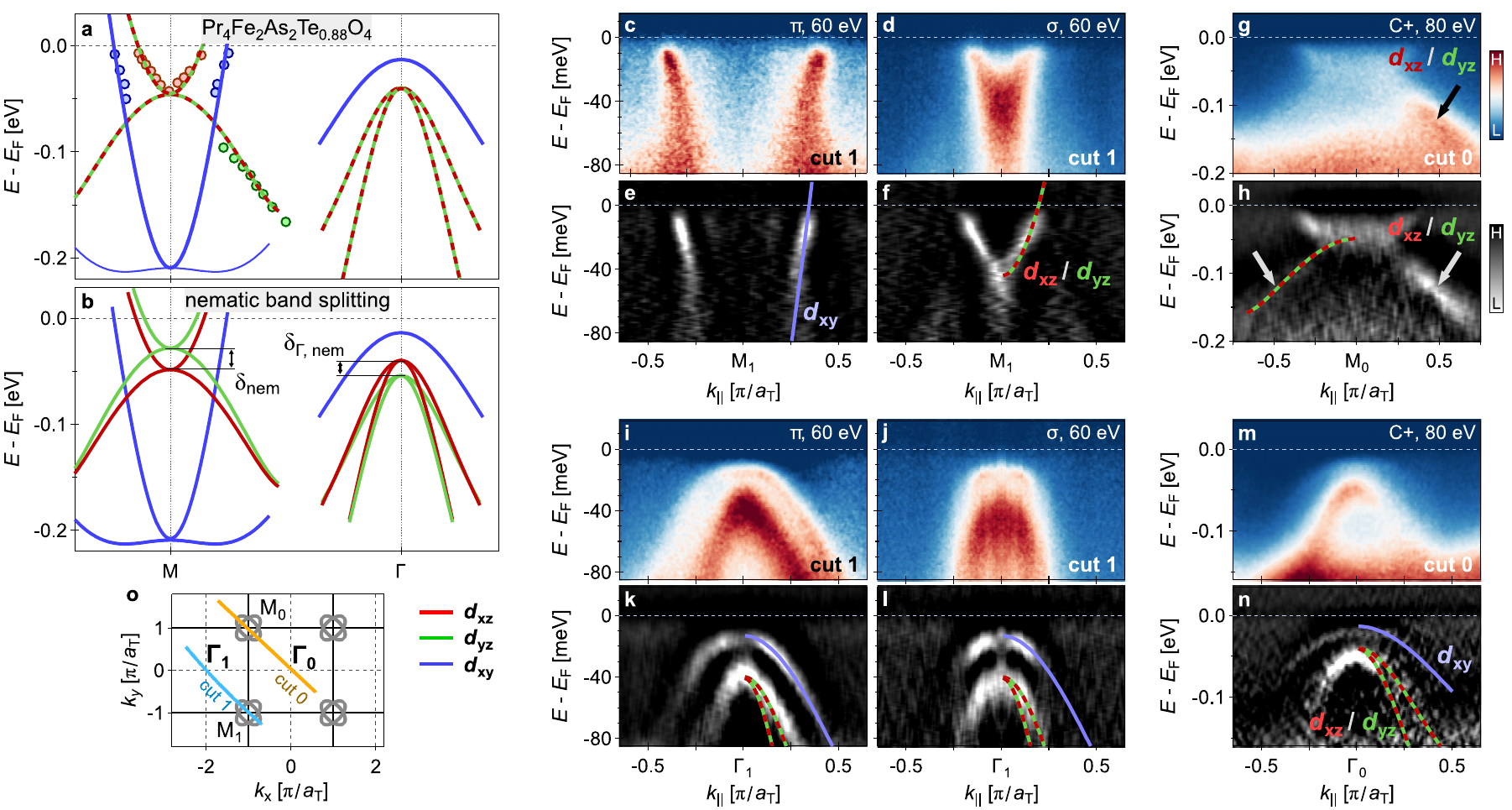}
 	\end{center}
 	\caption{\textbf{Absence of electronic nematicity in Pr$_{4}$Fe$_{2}$As$_2$Te$_{0.88}$O$_4$ (PFATO).}
 	\textbf{a,} Schematics of observed band dispersion in \PFATO. \textbf{b,} Schematics of band dispersion expected for the electronic nematic phase characterized by a finite nematic order parameter (energy splitting between \dxz ~and \dyz ~states) of $\delta_\mathrm{nem}\sim 20$~meV. 	
 	\textbf{c-h,} Diagonal cuts through M$_0$ and M$_1$, defined in panel (\textbf{o}), measured at 5.5 K with photon energy and polarisation as indicated, plotted in false-colour scale with `H' (`L') denoting high (low) intensity. \textbf{e,f,h,} Curvature plots of the raw spectra \textbf{(c,d,g)} calculated as described in Ref.~\onlinecite{zhangROSI2011}. All expected bands with \dxy, \dxz, and \dyz ~orbital character (blue, red and green lines) are identified. \textbf{i-n,} Similar spectra as \textbf{(c-h)} but along cuts through $\Gamma_0$ and $\Gamma_1$  as indicated in \textbf{(o)}. Spectra in \textbf{(j)} and \textbf{(l)} are symmetrised around $\Gamma_1$. \textbf{o,} Sketch of the Brillouin zone and electron-like Fermi surface sheets indicated in grey colours and cut 0  and cut 1 indicated in orange and blue.  Horizontal dashed  lines in \textbf{(a-n)} indicate the Fermi level \EF.
  	 }
	\label{fig:fig3}
 \end{figure*}

By analysing the structural peak splitting~\cite{McIntyrePRB1988}, we infer the orthorhombic order parameter $\delta=(a_\mathrm{O}-b_\mathrm{O})/(a_\mathrm{O}+b_\mathrm{O})$, where $a_\mathrm{O}$ and $b_\mathrm{O}$ are the orthorhombic lattice parameters (see Fig. \ref{fig:fig1} a). As shown in Fig.~\ref{fig:fig1} f, $\delta$ gradually increases and plateaus below $\sim50$ K at $12\times10^{-4}$. Importantly, $\delta$ retains its maximum value for all temperatures below \Tc\ (within error bars) indicating friendly coexistence with superconductivity (see inset of Fig. \ref{fig:fig1} f). This is in contrast to the strong phase competition between the electronically induced nematic phase of underdoped iron-pnictides in which nematicity gets suppressed below \Tc\ and C$_4$ rotational symmetry restores at low temperatures close to optimal doping  \cite{NandiPRL2010}. 
It has been suggested that the suppression of orthorhombicity in the electronic nematic phase is indirect and arises due to the competition for the same electronic states between magnetism and superconductivity\cite{RMFernandesNPHYS2014}. This suggests that orthorhombicity in \PFATO\ is not triggered by electronic degrees of freedom which compete with superconductivity.

In order to reveal rotational symmetry breaking in the electronic degrees of freedom, we performed high-resolution ARPES measurements on twinned samples. At this point it is worth noting that due to the very small sample size of $\leq 50 \times 50 \mu\mathrm{m}^2$ usual methods such as in-plane resistivity anisotropy on detwinned samples are not feasible. However, as ARPES directly accesses the electronic structure in momentum space it presents a suitable technique to measure the electronic nematic order parameter on twinned samples by resolving the \dxz\ -- \dyz\ orbital splitting \cite{yiPNAS2011}.

Figs.~\ref{fig:fig2} a-c show the Fermi surface and representative diagonal cuts through the tetragonal Brillouin zone (BZ) center ($\Gamma$) and corner (M). Vertical ($\sigma$) and horizontal ($\pi$) linear light polarisations are used to enhance bands of different orbital character (see supplementary Figs. S1 and S2). Furthermore, we use density functional theory (DFT) calculations to assign the orbital character of each measured band, as illustrated in Figs.~\ref{fig:fig2}~b and c and supplementary Fig. S1. Colours of the overlaid lines indicate the main orbital character of the DFT derived bands. Fig. \ref{fig:fig3} a summarizes the extracted electronic band structure along the $\Gamma$ -- M high symmetry cut. Due to the very large $c$-axis lattice constant we don't expect any significant out-of-plane dispersion, as suggested by our DFT calculations (see supplementary Fig. S1 a-b).

Electronic nematicity manifests as an energy splitting of the \dxz ~and \dyz ~orbitals, as sketched in Figs.~\ref{fig:fig1} b and ~\ref{fig:fig3} b. In the following, we disentangle the bands of different orbital character by using light polarisation analysis on a \textit{twinned} sample (Figs.~\ref{fig:fig3} c-n, supplementary Fig. S2-S7). Using $\pi$ and $\sigma$ polarised light we observe two electron-like bands around M$_1$, which we identify as the \dxy\ band (blue) and a superposition of the \dxz\ and \dyz\ bands (see Figs.~\ref{fig:fig3} c-f). Past reports on the electronically driven nematic phase in iron pnictides have shown that the nematic order parameter is largest at the M point\cite{yiNewJPhys2012}. Our curvature plots (Fig. \ref{fig:fig3} e,f) show that neither of the two bands splits in energy or momentum direction. A cut through M$_0$ with circular polarised light is shown in Fig. \ref{fig:fig3} g.
We observe the \dxz\ -- \dyz\ hole-like band at binding energy of $E_\mathrm{B} \sim -100$~meV. Similarly to the two electron-like bands, this band is also not energy-split within our experimental resolution (see Methods), as seen in the curvature and second derivative plots (Fig.~\ref{fig:fig3} h and supplementary Fig. S2). 
Figs.~\ref{fig:fig3} i-n depict the bands around the $\Gamma$ point in the first and second BZ, probed by different light polarisation. We observe the top of the three hole-like bands with each polarisation. 
The band closest to the Fermi level derives from the \dxy\ orbital, while the second and third band have \dxz\ -- \dyz\ character. For all three setup conditions, we observe that the \dxz\ and \dyz\ bands are degenerate at the $\Gamma$ point (see also supplementary Fig. S5).

In iron-pnictides, the structural orthorhombicity in the nematic phase is suppressed below \Tc\ (see Ref. \onlinecite{NandiPRL2010}), indicating a competition between superconductivity and electronic nematicity. Therefore, it is important to check if superconductivity is suppressing any possible band splitting in \PFATO. We recorded the temperature dependence of the bands at the M and $\Gamma$ point up to 56 K ($\sim 2 \times T_\mathrm{c}$, see supplementary Fig. S6 and S7) and observe that the \dxz\ -- \dyz\ bands remain degenerate up to highest measured temperatures at both high-symmetry points ($\Gamma$ and M).

We summarize our multifold high-resolution ARPES study on a twinned \PFATO\  sample in Fig.~\ref{fig:fig3} a. For comparison, Fig.~\ref{fig:fig3} b depicts the expected band structure of a twinned sample with a low-temperature electronic nematic phase (see supplementary Fig. S2 p), which is distinct from our observations. To further exclude that the here observed orthorhombic phase is driven by electronic degrees of freedom, we estimate the expected nematic order parameter ($\delta_\mathrm{nem}$) by comparing the low-temperature structural orthorhombic order parameter of \PFATO,  $\delta \sim 12\times10^{-4}$ to other iron-pnictides. We find that Ba(Fe$_{1-x}$Co$_x$)$_2$As$_2$, ($x\sim0.05$) shows a similar $\delta$ (see Ref.~\onlinecite{NandiPRL2010}) and therefore a band splitting of $\delta_\mathrm{nem} \sim 20$ meV (see Ref. \onlinecite{yiPNAS2011}) is not inconceivable. With the applied energy resolution ($< 5$ meV), such a splitting of the \dxz\ and \dyz\ states should be well detectable in our measurements. On the other hand, it has been suggested that $\delta_\mathrm{nem}$ generally scales with the orthorhombic transition temperature in the electronic nematic phase\cite{yinpjQuantMater2017}. This would set the expected band splitting in \PFATO\ to $> 100$ meV. Our experiments definitely exclude such a splitting.
The absence of the \dxz\ -- \dyz\ splitting (see Fig.~\ref{fig:fig3}) suggests that the structural orthorhombicity is not driven by electronic nematicity in \PFATO.

 \begin{figure}
 	\begin{center}
 		\includegraphics[width=1\columnwidth]{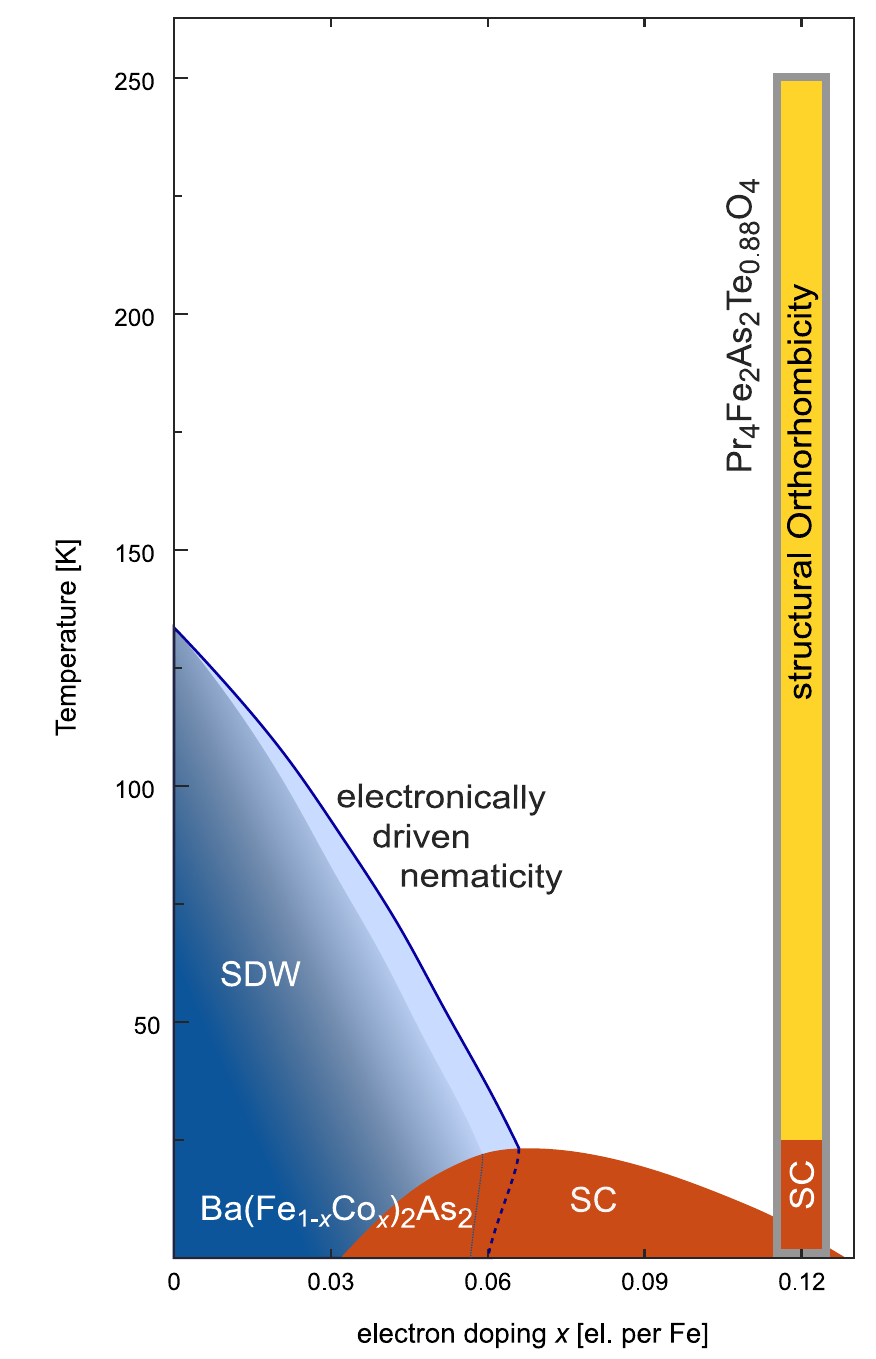}
 	\end{center}
 	\caption{\textbf{New iron-pnictide phase.}
 	Comparison between \PFATO\ and Ba(Fe$_{1-x}$Co$_{x}$)$_2$As$_2$. The nematic and spin-density wave (SDW) phase in underdoped Ba(Fe$_{1-x}$Co$_{x}$)$_2$As$_2$ follow a similar doping dependence and are intersecting the superconducting dome close to optimal electron doping at $\sim 0.06$ electrons per Fe. At roughly twice the doping ($x = 0.12$), \PFATO\ hosts a high-temperature onset of orthorhombicity at $T_\mathrm{ort} \sim 250$ K (100 K above $T_\mathrm{nem}$ of BaFe$_2$As$_2$), and a superconducting transition at $T_c \sim 25$ K (similar to optimally doped Ba(Fe$_{1-x}$Co$_{x}$)$_2$As$_2$). }
	\label{fig:fig4}
\end{figure}

In combination our XRD and ARPES study on the iron-based superconductor PFATO provide strong evidence for a new phase in the iron-pnictide phase diagram. The direct comparison between \PFATO\ and the widely studied Ba(Fe$_{1-x}$Co$_x$)$_2$As$_2$ system (Fig. \ref{fig:fig4}) highlights the difference between the electronic nematic phase and the here discussed structural orthorhombic phase from the perspective of the doping dependence. The phase transition of the electronic nematic phase has its highest $T_\mathrm{nem}$ at the parent compound which is then decaying once electrons are doped into the system. Furthermore, it is closely accompanied by a spin-density wave (SDW) phase which follows a similar doping-dependence. So far, a magnetic transition has not been found in \PFATO \cite{KatrychPRB2013, KatrychPRB2014, PisoniPRB2016}, but could possibly be stabilised under pressure, similar to bulk FeSe \cite{KothapalliNatComm2016}.  Furthermore, the absence of SDW order could be explained by the poorly fulfilled nesting condition, due to the absence of the hole-like Fermi surface around $\Gamma$ (see Fig. \ref{fig:fig2} a).  From this standpoint it is hard to imagine that the orthorhombic phase of 12 $\%$ electron-doped PFATO has a similar doping dependence and is of the same origin as the nematic phase. 

PFATO therefore presents an iron pnictide with a purely structural phase transition right on top of the superconducting dome. The structural phase transition is likely driven by phonon modes which make this system highly interesting for studying the influence of lattice vibrations on the  superconducting pairing mechanism. Our observation therefore calls for further experimental and theoretical investigations in order to understand the interplay between this structural instability and superconductivity. \\

\textbf{Acknowledgements:}
We are grateful to J\"org Schmalian for discussions.  
Acknowledgements goes to Diamond Light Source for time on beamline I05 under proposal SI16104 and thank all the beamline staff
for technical support. We acknowledge DESY (Hamburg, Germany), a member of the Helmholtz Association HGF, for the provision of experimental facilities. Parts of this research were carried out at P07 beamline at PETRA III synchrothron. C.E.M acknowledges support from the Swiss National Science Foundation under grant No. 200021-137783, P400P2$\_$183890, and P2EZP2$\_$175155. O.I., M.H., J.C., D.S, and J.C. acknowledge support from the Swiss National Science Foundation under grant No. BSSGIO$\_$155873, PP00P2$\_$150573 and through the Sinergia Network Mott Physics Beyond the Heisenberg Model.  \\

\textbf{Authors contributions:}
 SK and LF grew single crystals of Pr$_{4}$Fe$_{2}$As$_2$Te$_{0.88}$O$_4$. CEM, MH, DS, TKK, NX, JM, and JC carried out the ARPES experiments. ND, JC, OI, MvZ  and JC conducted the x-ray diffraction experiments. 
 CEM prepared the samples for both the ARPES and XRD experiments. MHF, QW, and JC supervised the project.
MS and NCP initiated the project. All authors contributed to the manuscript.  \\

\textbf{Methods:}
High-quality single crystals of Pr$_{4}$Fe$_{2}$As$_2$Te$_{0.88}$O$_4$ were synthesized by  high-pressure flux-growth. Previous X-ray diffraction measurements have shown that, at room temperature, PFATO crystallizes in a tetragonal \textit{I4/mmm} crystal structure (see Fig.~\ref{fig:fig1} c) with lattice parameters $a_T=4.02$ and $c=29.86$~\AA~\cite{KatrychPRB2014}. We selected the largest crystals -- $50\times50\times10$ $\mu$m$^3$ -- for ARPES and high-resolution hard x-ray diffraction (XRD) experiments. 
Our XRD experiment was carried out on the P07 beamline at PETRA III (DESY-Hamburg). A Pr$_{4}$Fe$_{2}$As$_2$Te$_{0.88}$O$_4$ crystal was mounted on the apex of a cactus needle to ensure sufficiently low background scattering. The sample was cooled inside a 10~T cryomagnet and oriented to access the $(h,k,0)$ scattering plane. Due to the small sample mass, the strongest measured Bragg reflection gave only 10 counts/ms using high-resolution monochromator settings.  \\
We performed ARPES experiments at both the SIS\cite{UFlechsigAIPCP2004} and I05 beamlines\cite{hoeschROSI2017a} -- Swiss Light Source, Switzerland and Diamond, UK respectively. Commercial Scienta analysers were used at both beamlines. Pristine surfaces were obtained through a top post cleaving method. Angle dependent x-ray photoelectron spectroscopy (XPS) measurements prove that the cleaving plane is the charge neutral Te layer (see supplementary Fig.~S10). Thermal and electrical grounding were realized through a combination of silver epoxy and graphite paste.  We estimated the Fermi level from  reference spectra of gold or copper in electrical and thermal contact with the PFATO sample. The energy resolution $\sigma_R=4.8$~meV (Gaussian standard deviation) was evaluated from the gold/copper references (see supplementary Fig. S8). Our momentum resolution was 0.1 degree or better. We performed density functional theory (DFT) band structure calculations of stoichiometric Pr$_{4}$Fe$_{2}$As$_2$TeO$_4$ using the WIEN2K package \cite{Blaha2001} (details are provided in the supplementary information).

\vspace{2mm}
\textbf{Data availability}\\
The data sets that support the findings in this study are available from the corresponding authors upon reasonable request.
\vspace{2mm}
 
\vspace{2mm}
\textbf{Competing interest}\\
The Authors declare no Competing Financial or Non-Financial Interests.
\vspace{2mm}

\textbf{Additional information}\\
Correspondence to: J.~Chang (johan.chang@physik.uzh.ch) and C.E.~Matt (cmatt@g.harvard.edu).
\vspace{4mm}


\end{document}